\documentclass[aps,prl,twocolumn,showpacs,groupedaddress]{revtex4}
\usepackage{epsfig}
\usepackage{longtable}

\begin{document}
\draft

\title{Modern Michelson-Morley experiment using cryogenic optical resonators}
\author{Holger M\"{u}ller$^{1,2}$, Sven Herrmann$^{1,2}$, Claus Braxmaier$^{2}$, Stephan Schiller$^3$, and Achim Peters$^1$}

\affiliation{$^1$Institut f\"{u}r
Physik, Humboldt-Universit\"{a}t zu Berlin, Hausvogteiplatz 5-7, 10117 Berlin, Germany,}
\email{holger.mueller@physik.hu-berlin.de}
\homepage{http://qom.physik.hu-berlin.de/}
\affiliation{$^2$Fachbereich Physik, Universit\"{a}t Konstanz,
78457 Konstanz, Germany,}
\affiliation{$^3$Institut f\"{u}r Experimentalphysik, Heinrich--Heine--Universit\"{a}t
D\"{u}sseldorf, 40225 D\"{u}sseldorf, Germany.}

\date{\today}

\begin{abstract}
We report on a new test of Lorentz invariance performed by comparing the resonance frequencies of
two orthogonal cryogenic optical resonators subject to Earth's rotation over $\sim 1$\,year. For a
possible anisotropy of the speed of light $c$, we obtain $\Delta_\theta c/c_0 = (2.6 \pm 1.7) \cdot
10^{-15}$. Within the Robertson-Mansouri-Sexl test theory, this implies an isotropy violation
parameter $\beta -\delta -\frac 12 = (-2.2\pm 1.5)\cdot 10^{-9}$, about three times lower than the
best previous result. Within the general extension of the standard model of particle physics, we
extract limits on 7 parameters at accuracies down to $10^{-15}$, improving the best previous result
by about two orders of magnitude.
\end{abstract}


\maketitle

Special relativity (SR) underlies all accepted theories of nature at the fundamental
level. Therefore, it has been and must be tested with ever increasing precision to
provide a firm basis for its future application. Such tests are also motivated by the
efforts to unify gravity with the other forces of nature, one of the outstanding open
challenges in modern science. In fact, many currently discussed models of quantum gravity
do violate the principles of SR. In string theory, for example, violation of Lorentz
invariance might be caused by spontaneous symmetry breaking \cite{KosteleckySSB}; in loop
gravity, effective low--energy equations (e.g., modified Maxwell equations
\cite{gambini}) that violate Lorentz invariance have been derived. Since the natural energy
scale for quantum gravity is the Planck scale, $E_p \sim 10^{19}$\,GeV, a direct
experimental investigation of quantum gravity effects is not feasible. It is, however,
possible to search for {\em residual} effects at an attainable energy scale in
experiments of outstanding precision. High-precision bounds on Lorentz violation might
give valuable hints for or against particular models of quantum gravity.

\begin{figure}[b]

\centering
\epsfig{file=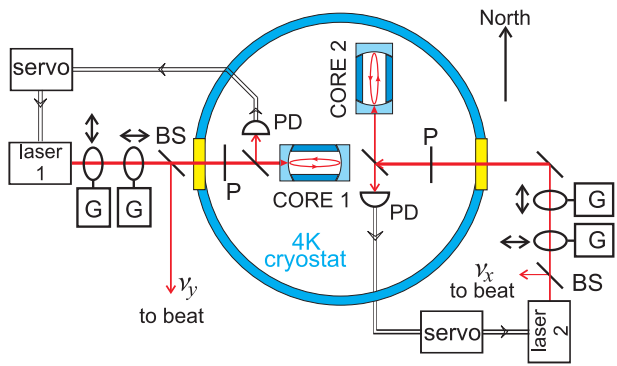, width=0.45\textwidth}
\caption{Setup. Inside a 4\,K cryostat, two COREs are located in a copper block to
provide common-mode rejection of thermal effects. LN2 is refilled automatically every
$\sim 3$\,h, LHe manually every $\sim 2$ days. Laser beams are coupled to the COREs via
windows, with polarizers P and lock detectors PD inside the cryostat. For active beam
positioning, beams pass through galvanomter (G) mounted glass plates. The horizontal and
vertical displacements are adjusted to maximize coupling into the cavities, as measured
using the $2f_m$ signal from the detector in reflection. Not shown is the setup for
measuring the frequency difference $\nu_x-\nu_y$ of the lasers, in which the beams are
overlapped on a high--speed photodetector and the beat frequency is measured against a
quartz oscillator stabilized to the GPS.
\label{setup}}
\end{figure}

A sensitive probe for Lorentz violation (in electrodynamics) is the Michelson-Morley (MM)
experiment \cite{MM}, that even predated the formulation of SR. It tests the isotropy of the speed
of light, one of SR's foundations. In the classic setup, one compares the speed of light $c$ in two
orthogonal interferometer arms by observing the interference fringes. If $c$ depends on the
direction of propagation, the fringes move if the setup is rotated (using, e.g., Earth's rotation
or a turntable). In 1979, Brillet and Hall \cite{BrilletHall} introduced the modern technique of
measuring a laser frequency stabilized to a resonance of an optical reference cavity. The frequency
of such a resonance is given by $\nu_{\rm cav}= m c/(2L)$, where $L$ denotes the cavity length and
$m=1,2, \ldots$ the mode number. Thus, a violation of the isotropy of $c$ can be detected by
rotating the cavity and looking for a resulting variation of $\nu_{\rm cav}$ by comparing the
frequency of the cavity--stabilized laser against a suitable reference. In our experiment (Fig.
\ref{setup}), we compare the frequencies $\nu_x$ and $\nu_y$ of {\em two} similar cavities
oriented in orthogonal directions, in analogy with the classical interferometer tests.
Compared to the single--cavity setup, this arrangement doubles the hypothetical signal
amplitude and provides some common--mode rejection of systematic effects.

Our results are obtained making use of the high dimensional stability of cryogenic
optical resonators (COREs). Constructed from crystalline sapphire, COREs show a low
thermal expansion coefficient ($10^{-10}/$K at 4.2\,K) and a remarkable absence of creep
(i.e., intrinsic length changes due to material relaxation); upper limits for the CORE
frequency drift are $<2$\,kHz/6\,months and $<0.1$\,Hz/h
\cite{COREs}, which makes COREs particularly well suited for high-precision measurements
such as relativity tests \cite{Braxmaier,MuellerASTROD}. For the MM experiment, it allows
us to rely solely on Earth's rotation. This avoids the systematic effects associated with
active rotation, which previous experiments had to use to overcome the creep of room
temperature resonators made from glass ceramics, e.g. ULE (ultra--low--expansion), on the
time--scale of a day.

A suitable theoretical framework for analyzing tests of Lorentz invariance is the general Standard
Model Extension (SME) \cite{Kostelecky,Kostelecky2002}. A Lagrangian formulation of the standard
model is extended by adding all possible observer Lorentz scalars that can be formed from known
particles and Lorentz tensors. In the Maxwell sector, the Lagrangian is $\mathcal L= \frac 14
F_{\mu \nu}F^{\mu\nu}+ \frac 14 (k_F)_{\kappa\lambda
\mu \nu}F^{\kappa \lambda} F^{\mu \nu}$ \cite{comment2}. The tensor $(k_F)_{\kappa
\lambda \mu \nu}$ (the greek indices run from $0 \ldots 3$) has 19 independent
components; they vanish, if SR is valid. 10 of its components, that can be arranged into
traceless symmetric $3\times 3$ matrices $(\tilde \kappa_{e+})^{AB}$ and $(\tilde
\kappa_{o-})^{AB}$, describe polarization--dependent effects. They are restricted to
$<2\cdot 10^{-32}$ by polarization measurements on light from astronomical sources
\cite{Kostelecky2002} and are assumed to be zero in the following. The remaining 9 components
describe boost invariance and isotropy of $c$ and can be arranged into traceless $3\times
3$ matrices $(\tilde \kappa_{e-})^{AB}$ (symmetric) and $(\tilde
\kappa_{o+})^{AB}$ (antisymmetric) plus one additional parameter.
They lead to a shift $\delta \nu$ of the resonance frequency of a cavity
\cite{Kostelecky2002} with a characteristic time signature and can thus be measured
in cavity experiments.\\

\begin{figure}[b]
\centering
\epsfig{file=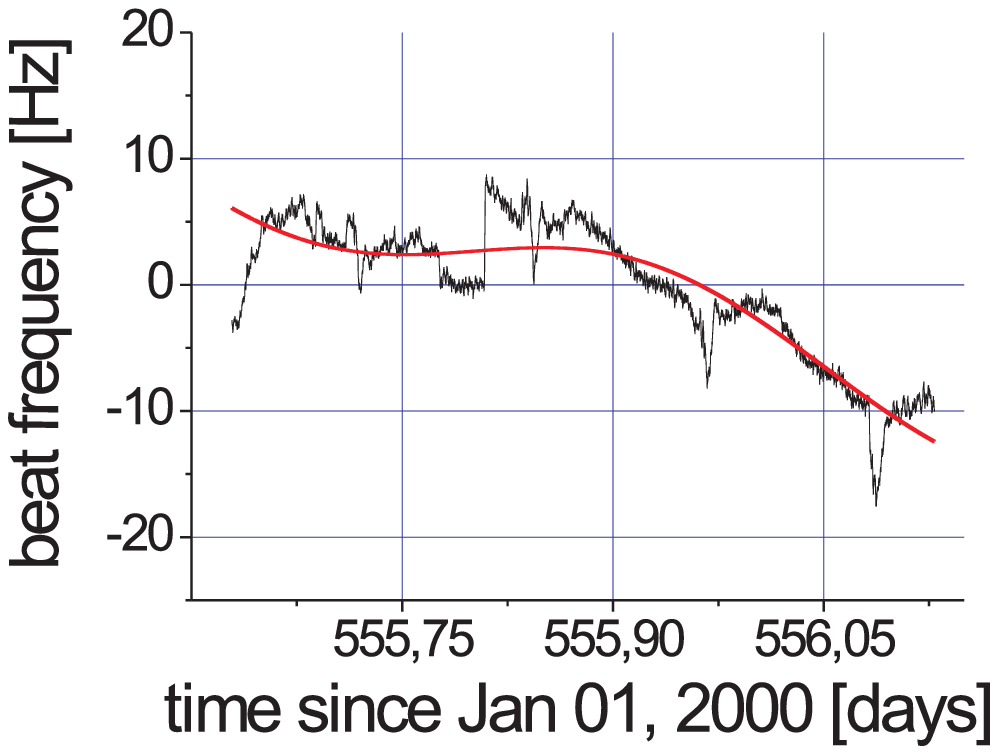, width=0.24\textwidth}
\epsfig{file=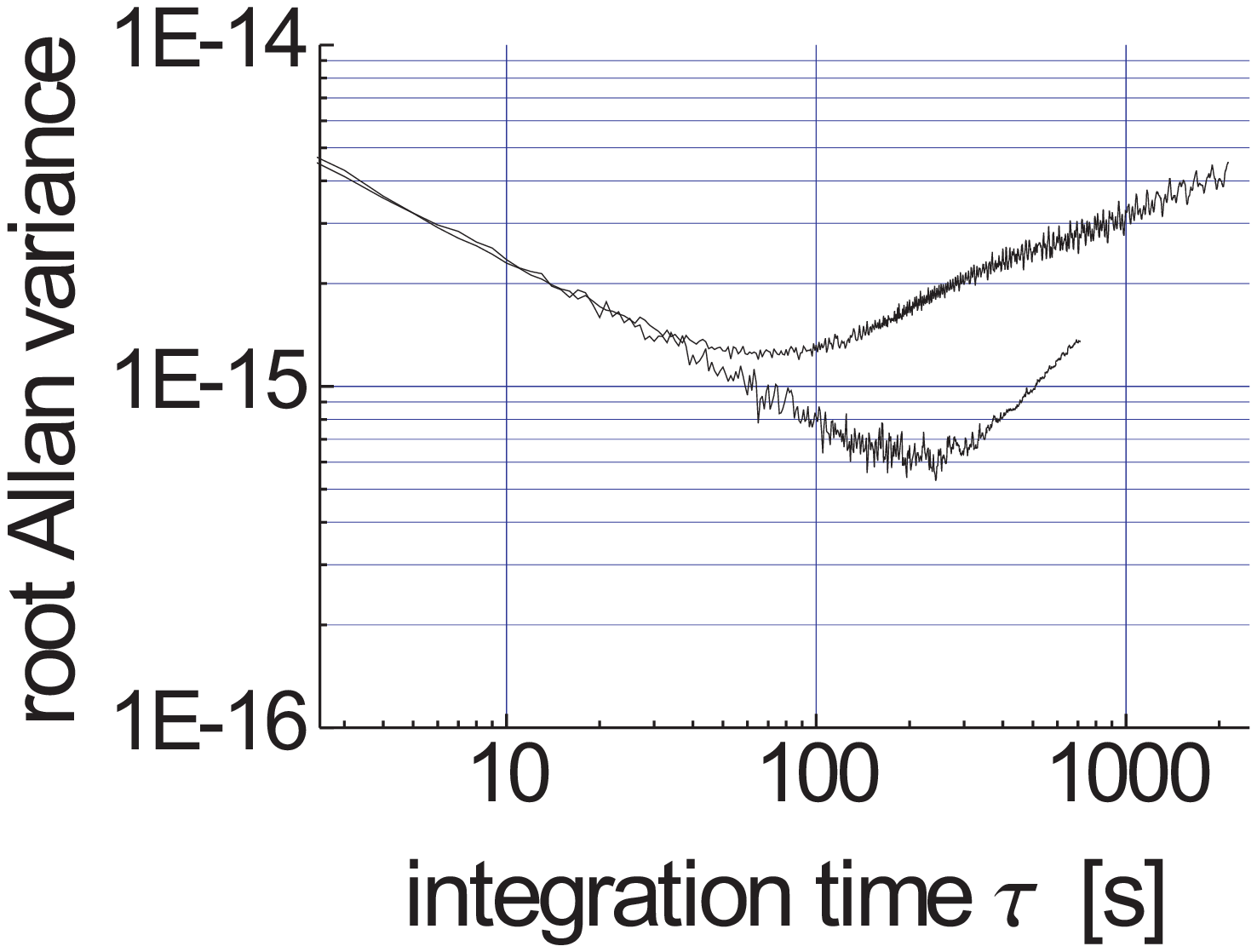, width=0.23\textwidth}
\caption{Left: Typical data set fitted with a 12\,h sinewave amplitude, a linear drift,
and a constant offset. Peaks occur every few hours due to automatic LN2 refills. Right: Root Allan
variance calculated from this data (upper curve), and from a quiet part between two LN2--refills
($118$ minutes starting at 555.87\,days; lower curve).
\label{datablock}}
\end{figure}


For our experiment (Fig. \ref{setup}), we use two $L=3$\,cm long COREs that feature
linewidths of 100\,kHz and 50\,kHz, respectively, (Finesses $\sim 10^5$), located inside
a liquid helium (LHe) cryostat with a liquid nitrogen (LN2) shield. Refills and the
evaporation of coolants cause mechanical deformations of the cryostat, which change the
resonator position. An automatic beam positioning system actively compensates for these
movements (Fig. \ref{setup}). The frequencies $\nu_{\rm las}$ of two diode-pumped Nd:YAG
lasers at 1064\,nm are stabilized (``locked") to resonances of the COREs using the
Pound-Drever-Hall method. The Nd:YAG laser crystal strain (generated by a piezo attached
to the crystal) and temperature were used for tuning the frequency. The phase modulation
at frequencies $f_m$ around 500\,kHz is also generated by crystal strain modulation using
mechanical resonances of the piezo. The light reflected from the COREs is detected inside
the cryostat (Fig. \ref{setup}) on Epitaxx 2000 InGaAs detectors. Down--conversion to DC
at $3f_m$ rather than $1f_m$ reduces the influence of residual amplitude modulation and
gives a higher signal-to-noise (S/N) ratio for the same circulating laser power inside
the cavity. For the detector signal, amplifiers consisting of 8 paralleled BF1009
dual-gate MOSFETs provide low current noise in spite of the high detector capacitance
(500\,pF). At $\sim 80$\,nW laser power impinging on the COREs, we achieve an error
signal S/N ratio of $\sim 1.5\cdot 10^4\cdot\sqrt{\tau /\mbox{s}}$. Between $80 \ldots
200$\,nW have been used, minimizing the change of $\nu_{\rm cav}$ due to laser heating of
the COREs ($\sim 10$\,Hz/$\mu$W). The error signal is generated using mini-circuits
ZHL-32 high-level amplifiers and SAY-1 23\,dBm double-balanced-mixers operating at $\sim
0$\,dBm RF amplitude and thus well below saturation. This provides highly linear
operation and proved very important for a low sensitivity to systematic disturbances.  On
a time scale of minutes, we reach a minimum relative frequency instability of the lasers
locked to the COREs of $7 \cdot 10^{-16}$, referred to a single CORE (Fig.
\ref{datablock}). Such a level is reached by the best ULE-cavity stabilized lasers
\cite{bergquist} only if a large linear drift $\sim 2$\,Hz/s is subtracted.

\begin{figure}[b]
\centering
\hspace{-0.4cm}
\epsfig{file=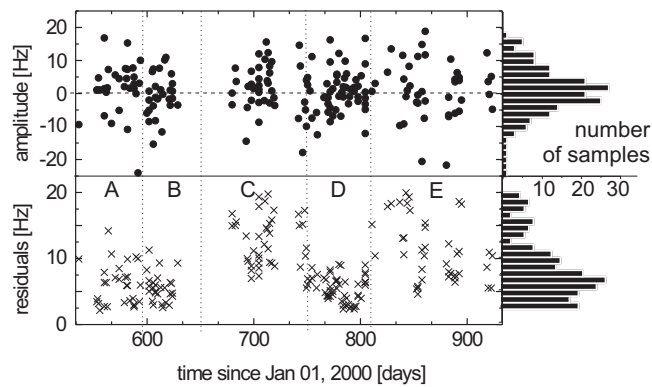, width=0.5\textwidth}
\caption{Fit results (upper part; dots) and fit residuals $\sqrt{\bar \chi^2}$
(lower part; crosses) for the amplitude of the cosine component of an assumed isotropy
violation signal at $2\omega_{\oplus}$. Strongly perturbed data sets are omitted when
$\sqrt{\bar \chi^2}>20$\,Hz (25 subsets). In 8 of these cases, a good fit was achieved
after discarding the first 1-3 hours (which likely contain residual perturbations from a
preceding LHe refill); these were then re--introduced to the analysis. The histograms
(bin size =1\,Hz) on the right hand side give the distribution of the fit results and
residuals.
\label{fitresults}}
\end{figure}

A number of systematic disturbances (e.g. residual amplitude modulation, parasitic
etalons, or mixer offset voltages) that cause lock--point shifts are compensated for
using a new technique: By an additional phase modulation of the laser beams, a part of
the laser power is shifted into side--bands, thus reducing the amplitude of the useful
part of the error signal. If the parasitic effects do not have sufficient frequency
selectivity to discriminate these sidebands against the carrier, modulating the amplitude
of the sidebands makes the lock--point shifts time--dependent, so they can be identified
and compensated for
\cite{offsetkomp}. This proved instrumental in achieving the
reliable and repeatable laser system performance required for accumulating enough data.


Except for a 10 day break around New Year 2002, the COREs were operated continuously at
4.2\,K over more than one year. Usable data (discounting data sets shorter than 12 hours
and data taken during adjustment or LHe refills) starts June 19, 2001 and was acquired
over 390\,days until July 13, 2002. A total of 146 data sets of 12\,h to 109\,h in
length, totaling 3461\,h, are available (Fig. \ref{datablock} and Fig.
\ref{fitresults}). 49 almost equally distributed data sets are longer then
24\,h. For extracting results, simultaneous least--squares fits with a constant offset, a
linear drift, and the amplitude of a sinusoidal signal at fixed frequency and phase as
suggested by the test theory are performed. Before fitting, the data sets are divided
into subsets of $12$\,h (or 24\,h for the 24\,h signals). This drops a fraction of the
data, but makes the resulting sinewave amplitudes independent of offsets in the data. We
obtain 199 fits of 12\,h data subsets (Fig. \ref{fitresults}).

The individual fit results (Fig. \ref{fitresults}) are combined for the final result by
coherent (vector) averaging. In the SME, the hypothetical Lorentz violation signal
$(\nu_x-\nu_y)/\bar \nu =\sum_{i} A^S_i \sin \omega_i T_\oplus + A^C_i \cos \omega_i
T_\oplus$, where $\bar \nu = (\nu_x+\nu_y)/2$, has Fourier components at 6 frequencies
$\omega_i$. As defined in
\cite{Kostelecky2002}, $T_\oplus = 0$ on March 20, 2001, 11:31 UT. The signal
components as given in Tab. \ref{sensitivity} are derived \cite{calculation} from Eq. (39) in
\cite{Kostelecky2002} (the cavity length $L$ is not substantially affected by the hypothetical
Lorentz violation \cite{TTSME}). Because our data extends over more than one year, in the vector
average we can resolve the 6 signal frequencies \cite{independence} and extract elements of the
traceless symmetric matrix
\[
\tilde \kappa_{e-} = \left( \begin{array}{ccc} a & 1.7 \pm 2.6 & -6.3\pm 12.4 \\
1.7 \pm 2.6 & b & 3.6 \pm 9.0 \\ -6.3 \pm 12.4 & 3.6 \pm 9.0 & -(a+b)
\end{array} \right) \cdot 10^{-15}
\]
with $a-b = 8.9 \pm 4.9$ (in the sun--centered celestial equatorial reference frame of
\cite{Kostelecky2002}). Likewise,
\[
\tilde \kappa_{o+}=\left(\begin{array}{ccc} 0 & 14 \pm 14 & -1.2 \pm 2.6 \\ -14 \pm 14 & 0 &
0.1 \pm 2.7 \\ 1.2 \pm 2.6 & -0.1 \pm 2.7 & 0 \end{array} \right)\cdot 10^{-11}
\]
for the antisymmetric matrix. One sigma errors are quoted. The limits on $\tilde
\kappa_{o+}$ are weaker because its elements enter the experiment suppressed by
$\beta_\oplus \sim 10^{-4}$, Earth's orbital velocity. All elements of $\tilde
\kappa_{o+}$ and all but one element of $\tilde \kappa_{e-}$ are obtained.
Compared to \cite{Lipa}, we improve the accuracy by about two orders of magnitude.
Moreover, while \cite{Lipa} give limits on linear combinations of the elements of $\tilde
\kappa_{o+}$, the present experiment allows individual determination, because of the
$>1$\,year span of our data.

\begin{table*}[t]
\caption{\label{sensitivity} Signal amplitudes $A^S_i, A^C_i$ from the SME, based on
\cite{Kostelecky2002}, and fit results. $\omega_\oplus \approx
2\pi/(23$h$ 56$min$)$ and $\Omega_\oplus = 2\pi/1$\,year denote the angular frequencies of Earth's
sidereal rotation and orbit; $\chi \approx 42.3^\circ$ is the colatitude of Konstanz, $\eta \approx
23.4^\circ$ the angle between the ecliptic and Earth's equatorial plane. The signal includes
contributions of order 1, or suppressed by either the Earth's orbital velocity $\beta_\oplus \sim
10^{-4}$ or by the velocity of the laboratory due to Earth's rotation $\beta_L \sim 10^{-6}$.  For
each $\omega_i$, we include only the largest term, in effect dropping all terms proportional to
$\beta_L$. A fit result that has been used to extract a parameter of the SME is set in bold face.
The unused fit results lead to additional (but weaker) limits on the elements of $\tilde
\kappa_{o+}$.}
\begin{center}
\begin{tabular}{|c|rr|rr|}
\hline $\omega_i$ & $A^S_i$ & Fit (Hz) & $A^C_i$ & Fit (Hz) \\ \hline
$\omega_\oplus-\Omega_\oplus$ & $\frac{\beta_\oplus}{2} \sin \chi \cos \chi [(\cos
\eta +1)(\tilde \kappa_{o+})^{XY} -\cos \eta (\tilde \kappa_{o+})^{XZ}] $ & ${\bf 1.82 \pm
1.91}$ & $\frac{\beta_\oplus}{2} \sin \chi \cos \chi \sin \eta (\tilde
\kappa_{o+})^{YZ}$ & $0.67 \pm 1.58$\\

$\omega_\oplus$ & $-\sin \chi \cos \chi (\tilde \kappa_{e-})^{YZ}$ & ${\bf -0.51
\pm 1.26}$ & $- \sin \chi \cos \chi (\tilde
\kappa_{e-})^{XZ}$ &  ${\bf 0.89 \pm 1.74}$\\

$\omega_\oplus+\Omega_\oplus$ & $\frac{\beta_\oplus}{2} \sin \chi \cos \chi [(\cos
\eta -1) (\tilde
\kappa_{o+})^{XY}-  \sin \eta (\tilde \kappa_{o+})^{XZ} ]$ & $-0.43 \pm 1.68$
& $\frac{\beta_\oplus}{2} \sin \chi \cos \chi \sin \eta (\tilde \kappa_{o+})^{YZ} $ &
$1.83
\pm 1.83$ \\

$2\omega_\oplus-\Omega_\oplus$ & $-\frac{\beta_\oplus}{4} (1+ \cos^2 \chi) (\cos
\eta +1)(\tilde
\kappa_{o+})^{YZ}$ & ${\bf -0.01 \pm 0.57}$ &  $-\frac{\beta_\oplus }{4} (1+ \cos^2 \chi)
(1+\cos \eta)(\tilde \kappa_{o+})^{XZ}$ & ${\bf 0.25 \pm 0.55}$ \\

$2\omega_\oplus$ & $ \frac 12 (1+ \cos^2 \chi) (\tilde \kappa_{e-})^{XY}$ & ${\bf 0.37
\pm 0.56}$ & $\frac 14 (1+ \cos^2 \chi)[(\tilde
\kappa_{e-})^{XX}-(\tilde
 \kappa_{e-})^{YY}]$ & ${\bf 0.97 \pm 0.53}$ \\

$2\omega_\oplus+\Omega_\oplus$ &  $\frac{\beta_\oplus}{4}(1+ \cos^2 \chi)(1-\cos
\eta)(\tilde
\kappa_{o+})^{YZ}$  & $0.50 \pm 0.55$ & $\frac{\beta_\oplus}{4} (1+ \cos^2 \chi)
(1-\cos \eta) (\tilde \kappa_{o+})^{XZ}$ & $0.75 \pm 0.58$ \\ \hline
\end{tabular}
\end{center}
\end{table*}

For comparison to previous work (e.g., \cite{BrilletHall,grieser,Braxmaier,Wolf}), we
also analyze our experiment within the Robertson-Mansouri-Sexl test framework
\cite{RobertsonMS}. It assumes generalized Lorentz transformations that contain
parameters $\alpha, \beta$, and $\delta$. In a preferred frame $\Sigma$ (usually the cosmic
microwave background), the speed of light $c_0$=const. In a frame $S$ moving with the velocity
$\vec v$ with respect to $\Sigma$, $c/c_0=1-(A+B\sin^2\theta) v^2/c_0^2$, where $\theta$ is the
angle between the direction of $c$ and $\vec v$. Both $A=\alpha -\beta +1$ and $B= \beta-\delta
-\frac 12$ vanish in SR. In this formalism, SR follows from (i) Kennedy-Thorndike (KT)-, (ii) MM-,
and (iii) Doppler shift experiments. The latter determine the time dilation coefficient $\alpha$
($=-\frac 12$ in SR) to $|\alpha + \frac 12| < 8 \cdot 10^{-7}$ \cite{grieser}. KT experiments test
velocity invariance of $c$, using the periodic modulation of $v$ provided by Earth's orbit
\cite{Braxmaier} or rotation \cite{Wolf}. Measuring the frequency
 of a cryogenic microwave cavity (that is proportional to $c$) against an H-maser,
\cite{Wolf} obtained $A=(-3.1\pm 6.9)\cdot 10^{-7}$.

While all three experiments are required for a complete verification of SR within this
framework, the MM experiment currently offers the highest resolution. In case of a
violation of isotropy, $B \neq 0$, for our experiment we obtain a periodic change of the
beat frequency $\nu_x-\nu_y$ at $2 \omega_\oplus$ \cite{RMSderivation}. For such a
signal, the experiment yields an amplitude $1.03 \pm 0.53 $\,Hz, or $B=(-3.1
\pm 1.6)\cdot 10^{-9}$. As the quality of the data is not uniform (Fig.
\ref{fitresults}), taking a weighted vector average is more appropriate here
\cite{independence}. We divide the data into the intervals A-E (Fig.\ref{fitresults})
with approximately uniform data quality within each. The averages over the intervals are
then combined to the final result, weighted according to their standard error. This leads
to a signal of $0.73 \pm 0.48$\,Hz, or
\[
B=\beta-\delta-\frac 12=(-2.2 \pm 1.5)\cdot 10^{-9} \, ,
\]
which we regard as the final result within the RMS framework. It has an inaccuracy about
three times lower than the best previous result $B=(3.0
\pm 4.9)\cdot 10^{-9}$\cite{BrilletHall,BHanalysis}.\\

In summary, we performed a modern Michelson-Morley experiment by comparing the
frequencies of two crossed cryogenic optical resonators subject to Earth's rotation over
a period spanning more than one year. Within the Robertson-Mansouri-Sexl framework, our
limit on the isotropy violation parameter is about three times lower than that from the
classic experiment of Brillet and Hall \cite{BrilletHall}. Moreover, we obtain limits on
seven parameters from the comprehensive extension of the standard model
\cite{Kostelecky,Kostelecky2002}, down to $10^{-15}$, about two orders of magnitude lower
than the only previous result \cite{Lipa}.\\ While the high long--term stability of COREs
allows one to use solely Earth's rotation, active rotation could still improve the
accuracy significantly. Due to the low drift of COREs compared to ULE cavities, the
optimum rotation rate would be relatively slow, which is desirable for minimizing the
systematics. At a rate of $\sim 0.2$/min ($\sim 5$/min was used in
\cite{BrilletHall}), one could utilize the optimum $\sim 7 \cdot 10^{-16}$ frequency stability
(Fig. \ref{datablock}) of the COREs, more than $10$ times better than on the 12\,h time
scale used so far. Since $\sim 500$ measurements (2 per turn) could be accumulated per
day, one should thus be able to reach the $10^{-17}$ level of accuracy. Further
improvements include fiber coupling and COREs of higher finesse. Finally, space based
missions with resonators are currently studied (OPTIS
\cite{OPTIS} and SUMO \cite{SUMO}).

We thank Claus L\"{a}mmerzahl for many valuable discussions and J\"{u}rgen Mlynek for making this
project possible. This work has been supported by the Deutsche Forschungsgemeinschaft and
the Optik-Zentrum Konstanz.


\begin{references}
\bibitem{independence}  In the SME analysis, unweighted averages are taken, since weighted
averages here would increase the interdependence of the fit results, typically $\sim
0.15$ and in all cases $\leq 0.36$. We can neglect it for the present purpose.
\bibitem{BHanalysis} The result is quoted here assuming $v=369$\,km/s \cite{lin96}
and taking into account a $\frac 14 [1+\cos(\chi_B)]^2\cos^2\xi$ factor for the
$\chi_B=50^\circ$ colatitude of the laboratory in Boulder.
\end{references}
\end{document}